\documentclass[prb,aps,twocolumn,amsmath,amssymb,floatfix,showpacs,
superscriptaddress]{revtex4}

\usepackage[dvips]{graphics}
\usepackage{color}
\definecolor{dred}{rgb}{0,0,0.7}

\begin{document}

\title{\textcolor{dred}{Curvature effect on spin polarization in a
three-terminal geometry in presence of Rashba spin-orbit interaction}}

\author{Santanu K. Maiti}

\email{santanu.maiti@isical.ac.in}

\affiliation{Physics and Applied Mathematics Unit, Indian Statistical
Institute, 203 Barrackpore Trunk Road, Kolkata-700 108, India}

\begin{abstract}

The robust effect of curvature on spin polarization is reported in a 
three-terminal bridge system where the bridging material is subjected to
Rashba spin-orbit interaction. The results are examined considering two 
different geometric configurations, ring- and linear-like, of the material
which is coupled to one input and two output leads. Our results exhibit
absolute zero spin polarization for the linear sample, while finite 
polarization is obtained in output leads for the ring-like sample.

\end{abstract}

\pacs{73.23.-b, 72.25.-b, 85.35.Ds, 71.70.Ej}

\maketitle

\section{Introduction}

The study of spin dependent transport in low-dimensional systems has
been largely dominated in the last few decades due to the rapid 
advancement in nano-scale science and technology~\cite{th1,th2,th3,th4,th5,
th6,th7,th8,san1,san2,san3,san4,th9,th10,san10,ex1,ex2,ex3}. Controlling 
electron's spin degree of freedom is extremely important for the development 
of quantum information processing as well as quantum computation~\cite{wolf}. 
The spin-orbit (SO) interaction which couples the electron's spin to the 
charge degree of freedom provides a much deeper insight for generating
spin current and also its manipulation~\cite{intrinsic1,intrinsic2,
intrinsic3,intrinsic4,intrinsic5,intrinsic6} rather than the usual 
methodologies~\cite{wang,xie}. Earlier, people were mainly using~\cite{wang,
xie} ferromagnetic leads or external magnetic field to get spin filtering 
action though these are not very suitable especially for low-dimensional 
systems, since in one case a large resistivity mismatch is observed while 
in the other case the main difficulty appears for confining a huge magnetic 
field into a narrow region, like a nano-ring.

Depending on the sources, spin-orbit interaction is classified in two
different categories: one is called extrinsic type which appears mainly
due to magnetic impurities, while the other is defined as intrinsic type 
that appears as a result of lacking of inversion symmetry. In this category 
generally two kinds of spin-orbit interactions are taken into account. They 
are called as Rashba and Dresselhaus SO interactions~\cite{rashba,dressel,
winkler}. The first one is associated with the inversion asymmetry of the 
structure and its strength can be regulated by means of external gate 
potential, and the second one is related to the bulk inversion asymmetry 
whose coupling strength depends on the material.

Considering the coupling of spin degree of freedom to the momentum of an 
electron, spin polarized currents in output terminals of a multi-terminal
conductor can be achieved from a purely unpolarized electron beam injected
to the input terminal~\cite{mlt1,mlt2,mlt3,mlt4,mlt5}. The existing 
literature suggest that a lot of theoretical progress has already been done 
to explore spin selective transmission through different model geometries. 
For example, a planar T-shaped conductor~\cite{mlt1} with a ring resonator 
exhibits polarized spin currents in outgoing leads in presence of Rashba 
SO interaction. In other work Peeters {\em et al.} have shown how a 
ring-like geometry can be utilized as an electron spin beam splitter 
exploring the possible quantum interference effect in presence of SO 
coupling~\cite{mlt3}. At the same time Nikolic and his group~\cite{mlt4,mlt5} 
put forward several key ideas in this particular field. 

In spite of the considerable volume of work available in this particular
area, a practically unexplored issue is how does the curvature of a
material which is clamped within input and output leads influence spin
polarization.
\begin{figure}[ht]
{\centering \resizebox*{6cm}{6cm}{\includegraphics{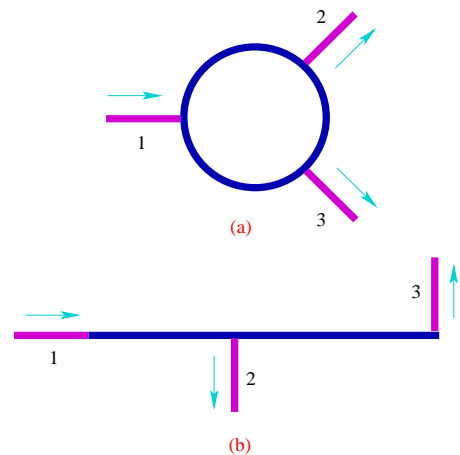}}\par}
\caption{(Color online). Schematic diagram of a three-terminal bridge 
setup, where the arrows correspond to the movement of electrons through
one input and two output terminals. Two different geometrical shapes, 
ring- and linear-like, of a particular sample subjected to Rashba SO 
interaction are taken into account to explore the curvature effect on spin
polarization in multi-terminal system.}
\label{model}
\end{figure}
To the best of our knowledge, this part is unaddressed so far. In the 
present work we essentially focus towards this direction. We investigate 
the curvature effect on spin polarization by considering two simple
geometries: a simple linear conductor and a ring-like geometry which is
formed by bending the chain.

The results are quite interesting. Using a tight-binding (TB) framework
and based on Green's function formalism we show that for a ring shaped 
conductor spin polarized currents are obtained in output leads of a 
multi-terminal geometry from a completely unpolarized beam of electrons, 
while absolute zero spin polarization is obtained for the linear conductor.

The rest of the paper is organized as follows. Section II illustrates two
different models together with a brief theoretical description to obtain 
spin polarization in two output leads. Section III contains numerical
results and discussion, and finally, in Section IV we summarize our essential
results.

\section{Model and theoretical framework}

In this section we describe two different systems of our study and present 
a general theory for calculating spin polarization coefficient $P^z$ in two
output leads based on Green's function formalism.

\subsection{Model and Hamiltonian}

The three-terminal bridge setup is schematically shown in Fig.~\ref{model},
where we take two different configurations of the same material. In one
configuration we choose a finite one-dimensional ($1$D) chain, which is 
then bent to form a $1$D ring to generate another configuration. In both 
these two cases the material, subjected to Rashba spin-orbit interaction,
is connected with one input (lead-1) and two output leads (lead-2 and lead-3).

A tight-binding framework is given under the nearest-neighbor hopping
approximation to describe the bridging material (ring/chain) and the 
side-attached leads. The TB Hamiltonian of the entire system reads as,
\begin{equation}
H = H_{\mbox{\tiny sample}} + H_{\mbox{\tiny leads}} + H_{\mbox{\tiny tun}} 
\label{eq1}
\end{equation}
where three different terms in the right side correspond to the Hamiltonians
of three different regions of the bridge system those are elaborately 
explained below.

The first term, $H_{\mbox{\tiny sample}}$, describes the Hamiltonian of the
conductor placed between the incoming and outgoing leads. Depending on its
geometry (ring-like or chain-like) the Hamiltonian looks different. For a
$N$-site linear conductor subjected to Rashba SO interaction the TB
Hamiltonian~\cite{liu,san9} gets the form:
\begin{eqnarray}
H_{\mbox{\tiny sample}} &=& \sum_n \mbox{\boldmath $c$}_n^{\dag} 
\mbox{\boldmath $\epsilon$} \mbox{\boldmath $c$}_n  
+ \sum_n \left(\mbox{\boldmath $c$}_n^{\dag} 
\mbox{\boldmath $t$} \mbox{\boldmath $c$}_{n+1} + h.c. \right) \nonumber \\
 & & + \sum_n \left(i\,t_{\mbox{\small so}} 
\mbox{\boldmath $c$}_n^{\dag} [\vec{\sigma} \times \hat{d}_{n,n+1}]_z 
\mbox{\boldmath $c$}_{n+1} + h.c. \right)
\label{eq2}
\end{eqnarray}
where, \\
$\mbox{\boldmath $c$}^{\dagger}_n=\left(\begin{array}{cc}
c_{n\uparrow}^{\dagger} & c_{n\downarrow}^{\dagger} 
\end{array}\right);$
$\mbox{\boldmath $c$}_n=\left(\begin{array}{c}
c_{n\uparrow} \\
c_{n\downarrow}\end{array}\right);$
$\mbox{\boldmath $\epsilon$}=\left(\begin{array}{cc}
\epsilon & 0 \\
0 & \epsilon \end{array}\right)$ \mbox{and} \\
$\mbox{\boldmath $t$}=t\left(\begin{array}{cc}
1 & 0 \\
0 & 1 \end{array}\right).$\\
~\\
In the above expression, $\mbox{\boldmath $c$}_{n\sigma}^{\dagger}$ and
$\mbox{\boldmath $c$}_{n\sigma}$ are the creation and annihilation operators,
respectively, for an electron with spin $\sigma(\uparrow,\downarrow)$ at 
the $n$-th atomic site of the sample. $\epsilon$ is the on-site energy and 
$t$ measures the isotropic nearest-neighbor hopping integral. The parameter
$t_{\mbox{\small so}}$ describes the Rashba SO coupling strength and the 
term $\vec{\sigma}$ gives the spin angular momentum of the electron. The
unit vector $\hat{d}_{n,n+1}$ describes the direction of the movement of an
electron between the sites $n$ and $n+1$.

When this $N$-site linear conductor is bent to form a ring the Hamiltonian
becomes~\cite{san5,san6,san7,san8},
\begin{eqnarray}
H_{\mbox{\tiny sample}} &=& \sum_i \mbox{\boldmath $c$}_n^{\dag} 
\mbox{\boldmath $\epsilon$} \mbox{\boldmath $c$}_n  
+ \sum_n \left(\mbox{\boldmath $c$}_n^{\dag} 
\mbox{\boldmath $t$} \mbox{\boldmath $c$}_{n+1} + h.c. \right) \nonumber \\
 & & + \sum_n \left(\mbox{\boldmath $c$}_n^{\dag} 
\mbox{\boldmath$t$}_{n,n+1} \mbox{\boldmath $c$}_{n+1} + h.c. \right)
\label{eq3}
\end{eqnarray}
where,\\
$\mbox{\boldmath $t$}_{n,n+1}=i t_{\mbox{\small so}} \left\{
\mbox{\boldmath$\sigma_x$} \cos\left(\frac{\varphi_n + 
\varphi_{n+1}}{2}\right) + \mbox{\boldmath$\sigma_y$} 
\sin\left(\frac{\varphi_n + \varphi_{n+1}}{2}\right)\right\}$ 
with $\varphi_n=2 \pi (n-1)/N$. All the other symbols used in Eq.~\ref{eq3} 
carry their usual meanings.

In our theoretical framework, three metallic leads are considered to be
identical, semi-infinite and free from any kind of impurities and spin-orbit
interaction. We can express them as,
\begin{equation}
H_{\mbox{\tiny leads}} = \sum_{\alpha} H_{\alpha} 
\label{eq4}
\end{equation} 
where $\alpha=1,2,3$ for the three leads. In the absence of any SO coupling 
$H_{\alpha}$ takes the form:
\begin{equation}
H_{\alpha} = \sum \limits_i \mbox{\boldmath$c$}_i^{\dagger} 
\mbox{\boldmath$\epsilon$}_l^{\alpha} \mbox{\boldmath$c$}_i + 
\sum_i \left(\mbox{\boldmath$c$}_i^{\dagger} \mbox{\boldmath$t$}_l^{\alpha}
\mbox{\boldmath$c$}_{i+1} + h.c. \right)
\label{eq5}
\end{equation}
with 
$\mbox{\boldmath $\epsilon$}_l^{\alpha}=\epsilon_l^{\alpha}
\left(\begin{array}{cc}
1 & 0 \\
0 & 1 \end{array}\right)$ and
$\mbox{\boldmath $t$}_l^{\alpha}=t_l^{\alpha}
\left(\begin{array}{cc}
1 & 0 \\
0 & 1 \end{array}\right)$.\\
~\\
where $\epsilon_l^{\alpha}$ and $t_l^{\alpha}$ are the site energy and 
nearest-neighbor hopping integral, respectively, in the $\alpha$-th lead. 
Other factors carry their usual meanings as stated earlier. Out of these 
three leads, lead-1 is treated as the input terminal, while the other two 
are considered as the output terminals and all of them are coupled to the 
conductor through the hopping integral $t_c$. Here we assume that the lead-1
is always attached to site $1$ and the other two leads are coupled to the 
sites $p$ and $q$, those are variables, of the conductor. Following the same 
footing as above, we can write the TB Hamiltonian to describe the 
conductor-to-lead coupling as,
\begin{equation}
H_{\mbox{\tiny tun}} = \sum_{\alpha} H_{{\mbox{\tiny tun}},\alpha}.
\label{eq6}
\end{equation}
Here,
\begin{equation}
H_{{\mbox{\tiny tun}}, \alpha} = \left[\mbox{\boldmath$c$}^{\dagger}_i
\mbox{\boldmath$t$}_c \mbox{\boldmath$c$}_n + \mbox{\boldmath$c$}^{\dagger}_n
\mbox{\boldmath$t$}_c \mbox{\boldmath$c$}_i \right]
\label{eq7}
\end{equation}
with 
$\mbox{\boldmath $t$}_c=t_c\left(\begin{array}{cc}
1 & 0 \\
0 & 1 \end{array}\right).$\\
The site $i$ corresponds to the boundary site of the lead, and it is coupled 
to the $n$-th site of the conductor, which is variable.

\subsection{Evaluation of polarization coefficient $P^z$ in terms of 
transmission probabilities}

The spin polarization coefficient in the output leads is defined 
as~\cite{san2,pola1,pola2,pola3},
\begin{equation}
P^z=\frac{T_{\uparrow \uparrow} + T_{\downarrow \uparrow} - 
T_{\uparrow \downarrow} - T_{\downarrow \downarrow}}{T_{\uparrow \uparrow} 
+ T_{\downarrow \uparrow} + T_{\uparrow \downarrow} 
+ T_{\downarrow \downarrow}}
\label{eq8}
\end{equation}
where, $T_{\sigma \sigma^{\prime}}$ gives the transmission probability
of an injecting electron with spin $\sigma$ which gets transmitted through 
the drain with spin $\sigma^{\prime}$. When $\sigma=\sigma^{\prime}$ we
get pure spin transmission, while for the other case spin flip transmission
is obtained. Equation~\ref{eq8} is the general expression of spin 
polarization coefficient between any two leads $i$ and $j$, and, for our 
three-terminal system we call the polarization coefficients in two outgoing
leads as $P_1^z$ and $P_2^z$. In the present approach we select the 
quantization direction along the $Z$ axis for simplification.

To calculate transmission coefficient $T_{\sigma \sigma^{\prime}}$ we use
Green's function formalism~\cite{datta1,datta2}. In this framework the 
two-terminal transmission
probability between the leads $i$ and $j$ is defined as 
$T_{\sigma \sigma^{\prime}}=\mbox{Tr}\left[\Gamma_i^{\sigma} G_c^r 
\Gamma_j^{\sigma^{\prime}} G_c^a\right]$. Here $G_c^r$ and $G_c^a$ are the
retarded and advanced Green's functions, respectively, of the sample
considering the effects of the electrodes. $G_c^r=\left(E-
H_{\mbox{\tiny sample}} - \sum\limits_{\sigma}\Sigma_1^{\sigma} -
\sum\limits_{\sigma}\Sigma_2^{\sigma} - \sum\limits_{\sigma}
\Sigma_3^{\sigma} \right)^{-1}$, where $E$ is the energy of an injecting 
electron, and $\Sigma_i^{\sigma}$'s ($i=1,2,3$) are the self-energies due 
to coupling of the conductor to the leads and $\Gamma_i^{\sigma}$'s are 
their imaginary parts. In Refs.~\cite{datta1,datta2} the detailed 
calculations of self-energy matrices are available.

\section{Numerical results and discussion}

Based on the above theoretical framework we now analyze our numerical 
results. Throughout the analysis we fix the electronic temperature of 
the system to absolute zero, and for simplification, we put $c=e=h=1$.
Other common parameters are as follows: $\epsilon=\epsilon_l^{\alpha}=0$
and $t=t_l^{\alpha}=t_c=1$. The Rashba SO coupling strength 
$t_{\mbox{\small so}}$ and all the energy scales are measured in unit 
of the hopping integral $t$.

Before focusing to the central point i.e., the curvature effect on spin
polarization in a multi-terminal (more than one output lead) system in
presence of Rashba SO interaction, we want to have a short glimpse on
the system where a conductor subjected to SO interaction is coupled to 
a single input and a single output lead i.e., a two-terminal system.
Following our extensive numerical calculations we can conclude that 
irrespective of the curvature of the bridging material, only SO 
interaction cannot induce spin polarization in output lead of a
two-terminal system. We verify it considering different geometrical 
shapes of the conductor, e.g., circle, square, triangle, polygon, linear,
etc. In few recent works~\cite{mlt1,moo} it has also been shown that 
only SO interaction
\begin{figure}[ht]
{\centering \resizebox*{8cm}{13cm}{\includegraphics{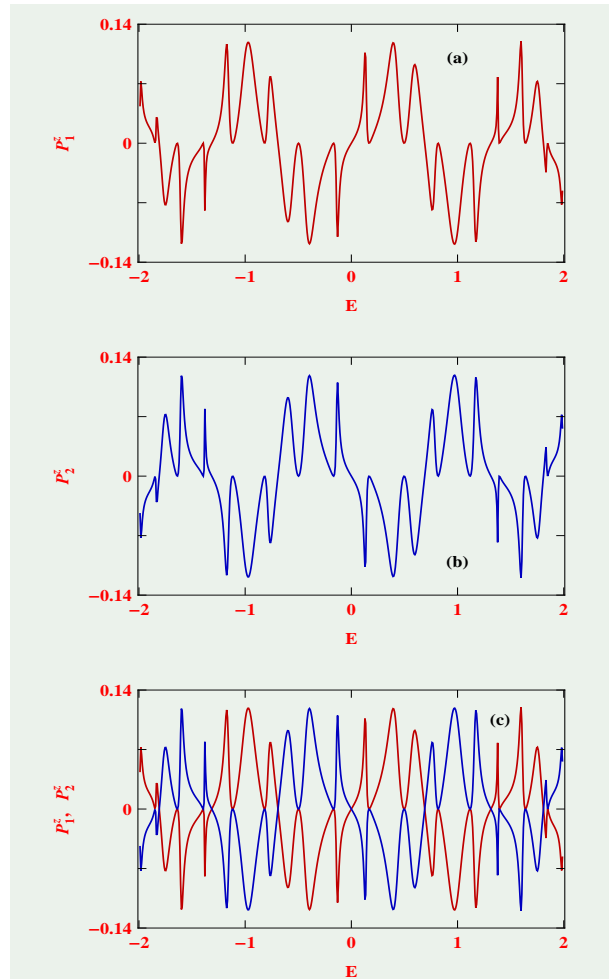}}\par}
\caption{(Color online). Energy dependence of spin 
polarization coefficients $P_1^z$ and $P_2^z$ in two output leads, in (a) 
and (b) respectively, for a three-terminal mesoscopic ring when the output 
leads are attached symmetrically with respect to the input lead. These 
coefficients ($P_1^z$ and $P_2^z$) are superposed with each other in (c) 
for comparison. The parameters are: $N=52$, $t_{\mbox{\small so}}=0.5$, 
$p=22$ and $q=32$.}
\label{ringpola}
\end{figure}
is incapable of producing spin polarization. The reason is that, in presence
of SO coupling the time-reversal symmetry is still preserved, and therefore,
it doesn't break the Kramer's degeneracy between the $|k\uparrow\rangle$ and 
$|-k\downarrow\rangle$ states which results vanishing spin current in the 
output lead of a two-terminal system. The degeneracy gets removed when the 
system is subjected to any kind of magnetic impurity or external magnetic
field. Under this situation a two-terminal system with SO coupling exhibits 
polarized spin currents~\cite{san2}. This phenomenon has already been 
established in the literature, but the essential issue of our present 
analysis -- the interplay between the curvature of the material and the 
multi-leads has not been addressed earlier.

To explore it, in Fig.~\ref{ringpola} we present the results for a 
three-terminal mesoscopic ring considering $N=52$ and 
$t_{\mbox{\small so}}=0.5$. Here, the two outgoing leads are attached
\begin{figure}[ht]
{\centering \resizebox*{8cm}{13cm}{\includegraphics{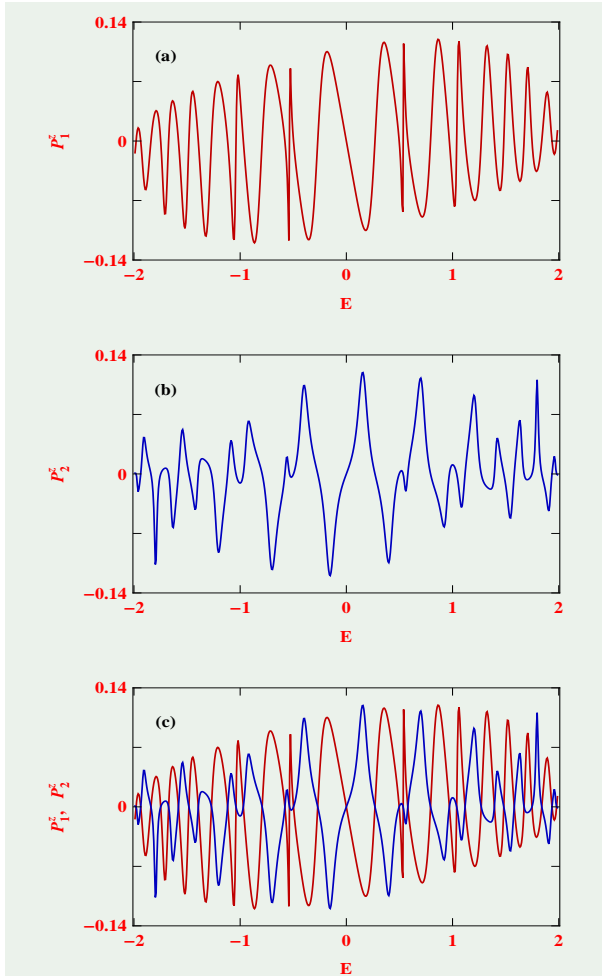}}\par}
\caption{(Color online). Energy dependence of spin 
polarization coefficients $P_1^z$ and $P_2^z$ in two output leads, in (a) 
and (b) respectively, for a three-terminal mesoscopic ring when the output 
leads are attached asymmetrically with respect to the source lead. These 
coefficients ($P_1^z$ and $P_2^z$) are superposed with each other in (c) 
for comparison. The parameters are: $N=52$, $t_{\mbox{\small so}}=0.5$, 
$p=27$ and $q=39$.}
\label{ringpolanew}
\end{figure}
symmetrically ($p=22$ and $q=32$) with respect to the incoming lead, as 
shown schematically in Fig.~\ref{model}(a). The upper panel of 
Fig.~\ref{ringpola} corresponds to the energy dependence of spin 
polarization coefficient for the one output lead, while for the other
output terminal it is shown in the middle panel of Fig.~\ref{ringpola}, 
and, finally they are placed together in the lower panel of this figure to 
compare the polarization coefficients properly. From these spectra it 
is observed that finite spin polarizations, associated with the energy
eigenvalues of the ring subjected to only SO interaction, are obtained 
in both the two outgoing leads though the system is free from any kind 
of magnetic impurities. Most interestingly, we also see that the 
coefficients $P_1^z$ and $P_2^z$ are exactly identical in magnitude and 
opposite in sign for each value of the incident electron energy $E$. 
This phenomenon can be explained as follows. The spin polarization 
\begin{figure}[ht]
{\centering \resizebox*{8cm}{9cm}{\includegraphics{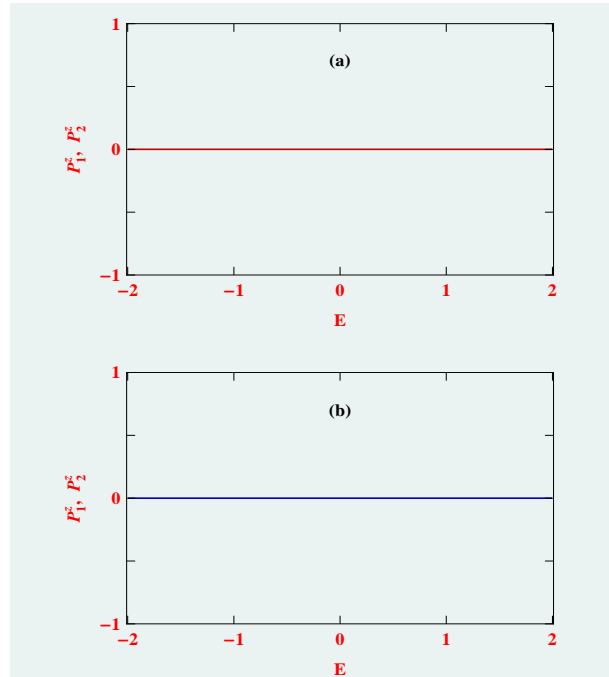}}\par}
\caption{(Color online). Energy dependence of spin 
polarization coefficients $P_1^z$ and $P_2^z$ in a three-terminal 
mesoscopic chain considering $N=52$ and $t_{\mbox{\small so}}=0.5$ for 
two different chain-to-lead interface geometries, where (a) $p=26$, $q=52$ 
and (b) $p=20$, $q=40$.}
\label{chainpola}
\end{figure}
coefficient $P^z$ describes the normalized difference among the up 
and down spin charge currents propagating through the outgoing leads, 
since in our present scheme we assume the quantized direction along the 
Z direction. In a multi-lead geometry i.e., when more than one outgoing 
lead is coupled to the system the Kramer's degeneracy between the 
$|k \uparrow \rangle$ and $|-k \downarrow \rangle$ gets removed and 
depending on the allowed paths of the moving electrons spin dependent
scattering takes place. The spin dependent force associated with the
SO coupling is responsible for this scattering. Now, in the ring-like
geometry electrons can go through two different paths, and accordingly,
the up and down spin electrons get deflected in two opposite directions 
by the spin dependent force during the movement of electrons through the 
ring geometry in presence of SO interaction which results spin selective 
transmission through the outgoing leads. This is the key aspect of 
observing mesoscopic spin Hall effect~\cite{san9} and the accumulation 
of opposite spin electrons on the opposite edges of a finite width conductor. 
Since the two outgoing leads are coupled symmetrically to the ring geometry
with respect to the incoming lead, $P_1^z$ and $P_2^z$ get equal magnitude.
Their sign reversals are also understood from Eq.~\ref{eq8}. 

By breaking the ring-lead interface geometry one can 
achieve spin polarizations with unequal magnitudes. The results are shown in
Fig.~\ref{ringpolanew} when the output leads are attached asymmetrically
with respect to the source lead. Here we fix $p=27$ and $q=39$, and all
the other parameters are kept identical as set in Fig.~\ref{ringpola}. It
is clearly seen that, unlike the symmetric configuration, the magnitudes of 
$P^z$ in two output leads are no longer equal for the asymmetrically 
connected ring-lead bridge setup. This is exclusively due to the quantum 
interference effect among the electronic waves propagating through 
different arms of the ring geometry. Apart from getting unequal magnitudes
of polarization coefficients in two output leads, all the other properties
remain exactly invariant as described earlier in the case of symmetric 
configuration.

From these results we can emphasize that, for the 
three-terminal ring geometry we get spin polarization due to SO interaction 
only, since the sample is free from any kind of magnetic impurity or external 
magnetic field, but a two-terminal ring geometry cannot provide polarize spin 
current under this situation. Here, it is important to note that although 
the Kramer's degeneracy between the $|k \uparrow \rangle$ and 
$|-k \downarrow \rangle$ gets removed by coupling the conductor with a third
lead or more, but it doesn't ensure to get non-vanishing spin polarization
in output leads which can be clearly understood from the following discussion.

The scenario becomes highly significant when the ring-like sample is
transformed into the linear-like one. The results are shown 
in Fig.~\ref{chainpola} considering two different chain-to-lead interface 
geometries. In one configuration (Fig.~\ref{chainpola}(a)) the outgoing 
leads are coupled to the sites $26$ and $52$ of the linear conductor, while 
these coupling sites are $20$ and $40$ for the other configuration 
(Fig.~\ref{chainpola}(b)).
The total number of atomic sites $N$ and the Rashba SO coupling are kept 
unchanged as taken in Fig.~\ref{ringpola}. From the spectra presented in 
Fig.~\ref{chainpola}, we interestingly see that both $P_1^z$ and $P_2^z$ 
drop exactly to zero for the entire energy band spectrum, and, these results 
are independent of the chain-to-lead interface geometry as well as the 
strength of the SO coupling, which we confirm through our considerable 
numerical work. This is really appealing in the sense that a same material 
subjected to SO coupling exhibits finite spin polarization for one 
geometrical shape, while absolute zero spin polarization is obtained in 
the other geometrical configuration for a multi-terminal bridge setup. 
This is solely due to the effect of curvature. For the linear chain the
up and down spin electrons can propagate only in a particular direction.
Either it can be along X or Y direction depending on the choice of the 
co-ordinate system. Under this situation, the spin dependent force which
essentially scatter opposite spin electrons becomes zero, and therefore,
no spin polarization is available even though the Kramer's degeneracy 
is broken for such a multi-terminal bridge setup. The disappearance of such
spin dependent force in a linear sample can be justified from the following 
analysis. For a linear chain the Rashba dependent Hamiltonian in the 
continuum model representation gets the form~\cite{xu}: 
$(t_{\mbox{\small so}}/\hbar) \sigma_y p_x$, assuming the movement of 
electrons along the X direction. Considering this Hamiltonian 
if we calculate \mbox{\boldmath$\ddot{x}$}, \mbox{\boldmath$x$} being
the position operator, then the output becomes exactly zero which 
immediately suggests vanishing spin dependent force since the force is
directly proportional to \mbox{\boldmath$\ddot{x}$}. To get a finite spin 
\begin{figure}[ht]
{\centering \resizebox*{8.25cm}{7cm}{\includegraphics{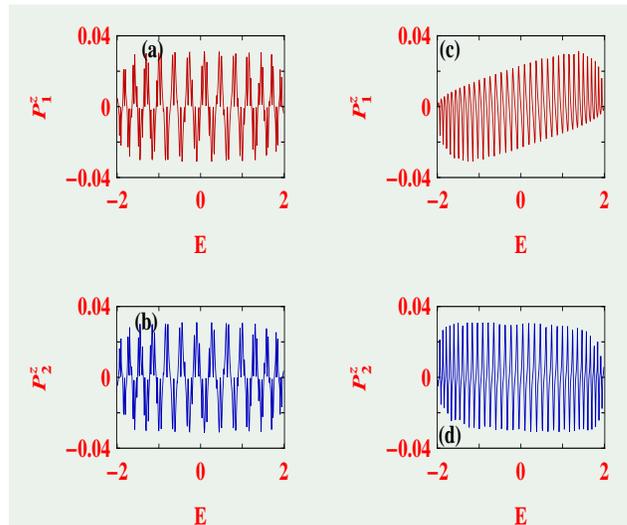}}\par}
\caption{(Color online). Energy dependence of spin 
polarization coefficients $P_1^z$ and $P_2^z$ in a three-terminal
mesoscopic ring for two distinct ring-to-lead interface geometries 
considering $N=200$ and $t_{\mbox{\small so}}=0.5$. The results for the
symmetric configuration are presented in the first column where we set
$p=82$ and $q=120$, while in the second column they are shown for the 
asymmetric configuration setting $p=101$ and $q=151$.}
\label{bigring}
\end{figure}
dependent force both the components $p_x$ and $p_y$ are needed in the Rashba 
term which is not possible in the case of a linear chain. Therefore, spin 
dependent scattering is no longer available, even in the presence of 
multi-leads. Now, an important point which should be noted here that when 
a linear conductor is properly bent to form a regular ring shaped geometry, 
the SO coupling strength may be affected due to its curvature, but that 
doesn't at all change the present physical scenario, and therefore, we 
consider the identical coupling strength for both these two geometrical 
configurations for the sake of simplification in our model calculations.

\begin{figure}[ht]
{\centering \resizebox*{8cm}{9cm}{\includegraphics{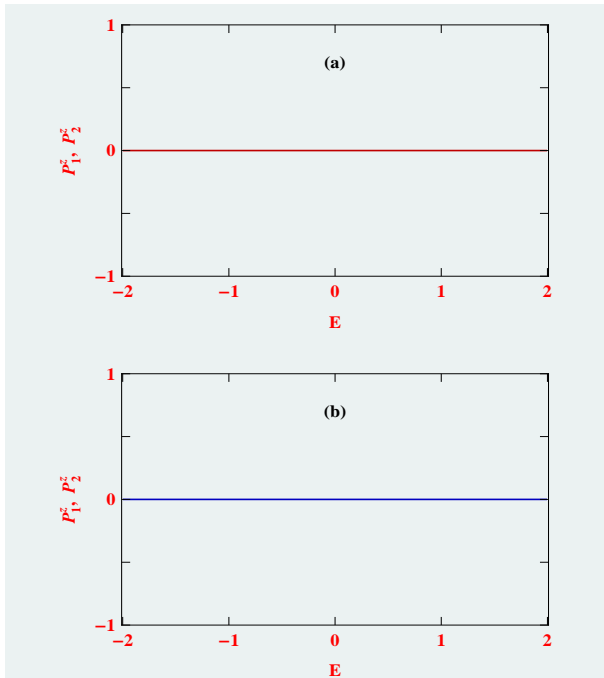}}\par}
\caption{(Color online). Energy dependence of spin 
polarization coefficients $P_1^z$ and $P_2^z$ in a three-terminal 
mesoscopic chain considering $N=200$ and $t_{\mbox{\small so}}=0.5$ for 
two different chain-to-lead interface geometries, where (a) $p=100$, $q=200$ 
and (b) $p=50$, $q=150$.}
\label{bigchain}
\end{figure}

The results presented so far to explore the curvature effect
on spin polarization in a three terminal geometry are computed for a
conductor with only $52$ sites. Keeping in mind a possible experimental 
realization one may think how such a small sized conductor can be used to 
design a conductor-lead bridge setup. To establish this fact now we present 
the spin polarization coefficients $P_1^z$ and $P_2^z$ taking a 
$200$-site conductor for its two different shapes as considered earlier
i.e., ring-like and the linear-like one. For the ring-like geometry, the
results are presented in Fig.~\ref{bigring}, whereas for the other case
they are shown in Fig.~\ref{bigchain}, and for both these two cases the
results are computed for two distinct electrode-to-conductor configurations
(symmetric and asymmetric) to justify the robustness of our investigation.
From the spectra given in Fig.~\ref{bigring} it is observed that, like
Fig.~\ref{ringpola}, here also the polarization coefficients $P_1^z$ and 
$P_2^z$ become exactly identical in magnitude and opposite in sign (1st 
column of Fig.~\ref{bigring}) when the ring is attached symmetrically to 
the outgoing leads. These coefficients $P_1^z$ and $P_2^z$ are no longer 
identical in magnitude for the asymmetric ring-lead configuration (2nd 
column of Fig.~\ref{bigring}), as expected. This is exactly what we get 
in a $52$-site ring as shown in Fig.~\ref{ringpolanew}. For the linear-shaped
conductor with $N=200$ a vanishing spin polarization is obtained
(Fig.~\ref{bigchain}) in its output leads for the entire energy band region
irrespective of the chain-to-lead interface geometry and it is exactly 
similar in nature which we get earlier in the case of a $52$-site chain 
(Fig.~\ref{chainpola}). From these results we can emphasize that apart from 
getting more peaks and dips in $P^z$-$E$ spectrum for the ring-shaped 
conductor with increasing $N$, all the basic physical properties i.e.,
non-vanishing spin polarization coefficients with equal and/or unequal
magnitudes associated with the ring-lead interface geometry and absolute zero 
spin polarization in the linear-like conductor irrespective of its coupling
configuration to side-attached leads remain exactly unchanged. For a very
large sized ring several peaks and dips appear in $P^z$-$E$ spectrum from
which it may seem that the spectrum is quasi-continuous, but looking it 
carefully one can always find distinct peaks since $N$ is finite. Thus,
our essential goal of getting finite and vanishing spin polarizations
in a multi-terminal geometry constructed with the same material with 
different curvatures is established.

\begin{acknowledgments}
The author is thankful to Prof. S. Sil and M. Dey for useful discussions.
\end{acknowledgments}

\section{Conclusion}

To summarize, in the present communication we establish the curvature effect
on Rashba spin-orbit interaction induced spin polarization in a three-terminal
bridge setup within a tight-binding framework based on Green's function 
formalism. The results are analyzed considering two different shaped 
geometries of the same material. In one configuration we select the bridging
material in a linear-like, which is then bent to form a ring-like geometry.
Quite interestingly, we find that finite polarization is obtained in two 
output leads for ring shaped geometry, while absolute zero spin polarization
is noticed when the sample becomes linear. This phenomenon also holds
true even for any other higher-terminal bridge setup and independent of the
lead-conductor interface geometries.

In addition to this ring-like geometry one might expect finite spin 
polarization in output leads for other geometrical configurations, except 
the linear one, which essentially leads to the robust effect of curvature 
on spin polarization in a multi-terminal bridge system.

All the results described in this communication are worked out at absolute 
zero temperature, though the finite temperature extension of this analysis 
is extremely trivial. The thing is that at finite temperature no new 
phenomenon will appear and all the physical pictures presented here remain 
unaltered even at finite (low) temperature since the broadening of energy 
levels of the conductor due to its coupling with the side-attached leads 
is too large compared to the thermal broadening~\cite{datta1,datta2,san11,
san12,san13}.

In the present work, we ignore the effect of on-site electron-electron (e-e)
interaction. We can incorporate this effect in our formalism in different
ways. One possible route is the mean field approximation~\cite{ee1,ee2,ee3,
ee4,ee5}. But, for this particular study e-e interaction doesn't provide 
any such new insight since it cannot scatter up and down spin electrons 
in opposite edges of the sample, as spin-orbit interaction does. Only some 
modifications in magnitudes of $P_1^z$ and $P_2^z$ can be expected.

Before we end, it should be noted that to explore the 
effect of curvature on spin polarization in a three-terminal bridge setup 
we compute our numerical results considering some typical values of the
parameters describing the systems. But, all these physical properties i.e., 
vanishing spin polarization in output leads of a linear-like conductor and 
finite spin polarization for a ring-like geometry remain absolute unchanged
for any other set of parameter values. These features certainly demand
an experiment in this line.

\end{document}